\documentclass[epj,nopacs]{svjour}

\usepackage[dvips]{graphics}
\usepackage{graphicx}
\usepackage{cite}

\newcommand{\epem}     {\ensuremath{\mathrm{e^+e^-}}}
\newcommand{\roots}    {\ensuremath{\sqrt{s}}}

\newcommand{\znull}    {\ensuremath{\mathrm{Z^0}}}
\newcommand{\mz}       {\ensuremath{m_{\znull}}}
\newcommand{\as}       {\ensuremath{\alpha_{\mathrm{S}}}}
\newcommand{\ash}      {\ensuremath{\hat{\alpha}_{\mathrm{S}}}}
\newcommand{\asmz}     {\ensuremath{\as(\mz)}}

\newcommand{\stat}     {\ensuremath{\mathrm{(stat.)}}}

\newcommand{\expt}     {\ensuremath{\mathrm{(exp.)}}}
\newcommand{\had}      {\ensuremath{\mathrm{(had.)}}}
\newcommand{\theo}     {\ensuremath{\mathrm{(theo.)}}}
\newcommand{\xmu}      {\ensuremath{x_{\mu}}}
\newcommand{\bt}       {\ensuremath{B_{\mathrm{T}}}}
\newcommand{\bw}       {\ensuremath{B_{\mathrm{W}}}}
\newcommand{\mh}       {\ensuremath{M_{\mathrm{H}}}}

\newcommand{\thr}      {\ensuremath{1-T}}
\newcommand{\cp}       {\ensuremath{C}}

\newcommand{\ytwothree}{\ensuremath{y_{23}^\mathrm{D}}}

\newcommand{\oaaa}     {\ensuremath{\mathcal{O}(\as^3)}}
\newcommand{\chisq}  {\ensuremath{\chi^2}}
\newcommand{\chisqd} {\ensuremath{\chi^2/\mathrm{d.o.f.}}}
\newcommand{\bbbar}  {\ensuremath{\mathrm{b\overline{b}}}}
\newcommand{\qqbar}  {\ensuremath{\mathrm{q\overline{q}}}}

\begin{document}

%\title{Determination of the Strong Coupling \boldmath{\as} from 
%hadronic Event Shapes and NNLO QCD predictions using JADE Data}

\title{Determination of the Strong Coupling \boldmath{\as} from
hadronic Event Shapes with ${\cal O}(\as^3)$ and resummed QCD
predictions using JADE Data}

\titlerunning{Determination of the Strong Coupling \boldmath{\as} $\ldots$}

\author{S. Bethke\inst{1} \and S. Kluth\inst{1} \and 
        C. Pahl\inst{1}\inst{2} \and J. Schieck\inst{1} \and the 
        JADE Collaboration\thanks{The members of the JADE collaboration
        are listed in~\cite{authors}}}

\authorrunning{S. Bethke et al.}

\institute{Max-Planck-Institut f\"ur Physik, F\"ohringer Ring 6,
D-80805 M\"unchen, Germany \and
Excellence Cluster Universe, Technische Universit\"at
M\"unchen, Boltzmannstr. 2, D-85748 Garching, Germany}

\date{Received: date / Revised version: date}

\abstract{Event Shape Data from \epem\ annihilation into hadrons
collected by the JADE experiment at centre-of-mass energies between
14~GeV and 44~GeV are used to determine the strong coupling \as.  QCD
predictions complete to next-to-next-to-leading order (NNLO),
alternatively combined with resummed next-to-leading-log-approximation
(NNLO+NLLA) calculations, are used.  The combined value from six
different event shape observables at the six JADE centre-of-mass
energies using the NNLO calculations is
\begin{displaymath} 
\asmz= 0.1210\pm 0.0007\stat\pm 0.0021\expt\pm 0.0044\had\pm 0.0036\theo
\end{displaymath} 
and with the NNLO+NLLA calculations the combined value is 
\begin{displaymath}
\asmz= 0.1172\pm 0.0006\stat\pm 0.0020\expt\pm 0.0035\had\pm 0.0030\theo \;\;.
\end{displaymath}
The stability of the NNLO and NNLO+NLLA results with respect to
missing higher order contributions, studied by variations of the
renormalisation scale, is improved compared to previous results
obtained with NLO+NLLA or with NLO predictions only. The observed
energy dependence of \as\ agrees with the QCD prediction of asymptotic
freedom and excludes absence of running with 99\% confidence level.}

\maketitle

\section{Introduction}

Analyses of events originating from \epem\ annihilation into hadrons
allow studies~\cite{biebel01a,dasgupta03,dissertori03,kluth06} of
Quantum Chromodynamics (QCD), the theory of the strong
interaction~\cite{fritzsch73,gross73a,gross73b,politzer73}.
Comparison of observables like jet production rates or event shapes
with theoretical predictions provides access to the determination of
the strong coupling \as.  Recently significant progress in the
theoretical calculations of event shape observables has been made and
next-to-next-to-leading order (NNLO) predictions are now
available~\cite{gehrmannderidder07b} as well as matching with resummed
calculations in the next-to-leading-log-approximation
(NLLA)~\cite{gehrmann08}.  As a first application, measurements of
\as\ at centre-of-mass-system (cms) energies between $\roots=91$~GeV
and $\roots=206$~GeV were presented~\cite{dissertori07}.  The same
theoretical predictions are used in this paper to determine the strong
coupling \as\ from JADE\footnote{JApan-Deutschland-England} data
recorded at lower cms energies.  As in~\cite{dissertori07} and the
previous standard LEP and JADE analyses~\cite{kluth06} we use Monte
Carlo simulations to treat hadronisation effects.  In \cite{davison08}
data for thrust at cms energies $14 \le \roots \le 207$~GeV are
analysed with combined NNLO+NLLA calculations and an analytic model
for non-perturbative physics.

The JADE experiment operated at the
PETRA\footnote{Positron-Elektron-Tandem-Ring-Anlage} \epem\ collider
at DESY\footnote{Deutsches Elektronen SYnchrotron}, Hamburg, Germany.
The data taken in the years from 1979 to 1986 cover cms energies
between 12 and 46.4~GeV.

\section{The JADE Detector}

A detailed description of the JADE detector can be found
in~\cite{naroska87}.  For this analysis, tracks from charged particles
and energy deposits in the electromagnetic calorimeter are used. The
main tracking device, a large volume jet chamber was located in a 0.48
T solenoidal magnetic field.  The electromagnetic calorimeter
consisted of 2520 lead glass blocks in the barrel and 192 lead glass
blocks in both endcaps with radiation length varying between 9.6 in
the endcaps and up to 15.7 in the barrel.

\section{Data and Monte Carlo Samples}

The Data and Monte Carlos samples utilised in this analysis are
identical to those used in a previous determination of
\as~\cite{jader4}.  The data correspond to a total integrated
luminosity of 195/pb taken at average cms energies between 14.0~GeV
and 43.8~GeV. The breakdown of the data samples, including the average
cms energies, the energy ranges, data taking periods, integrated
luminosities and the overall numbers of selected hadronic events are
summarised in table~\ref{TableNumberEvents}.

\begin{table}[htb!]
\caption{ The average centre-of-mass energy \roots, 
the energy range, data taking period, collected integrated luminosity
$L$ together with the number of selected data events.}
\label{TableNumberEvents}
\begin{tabular}{ccrcr}
\hline\noalign{\smallskip}
\roots & energy       & year & $L$  &  selected  \\
GeV & range in GeV &      & (1/pb)    &  events  \\
\noalign{\smallskip}\hline\noalign{\smallskip}
14.0 & 13.0--15.0 & 1981       & 1.46 &  1783 \\
22.0 & 21.0--23.0 & 1981       & 2.41 &  1403 \\
34.6 & 33.8--36.0 & 1981--1982 & 61.7 &  14313 \\
35.0 & 34.0--36.0 & 1986       & 92.3 &  20876 \\
38.3 & 37.3--39.3 & 1985       & 8.28 &  1585 \\
43.8 & 43.4--46.4 & 1984--1985 & 28.8 &  4374 \\
\noalign{\smallskip}\hline
\end{tabular}
\end{table}

Monte Carlo events are generated in large numbers to correct the data
for experimental acceptance, resolution effects and background. Events
are simulated using PYTHIA~5.7~\cite{jetset3}, and for systematic
studies with HERWIG~5.9~\cite{herwig}.  Subsequently the generated
events are processed through a full simulation of the JADE detector
and are reconstructed in the same way and with the same program chain
as the data.  For comparison with the corrected data and for the
correction of hadronisation effects large samples of Monte Carlo
events have been produced using PYTHIA~6.158, HERWIG~6.2 and
ARIADNE~4.11~\cite{ariadne3}.  We use the model parameters as
determined at \roots=91~GeV by the OPAL experiment at the LEP \epem\
collider~\cite{OPALPR141,OPALPR379}.

\section{Experimental Procedure}

\subsection{Event Selection}

The selection of identified and well measured hadronic event
candidates follows the procedure outlined in~\cite{jader4}.  Events
with a large momentum imbalance due to photons emitted in the initial
state are rejected.  The event selection is based on minimal
requirements for charged particle multiplicity, visible energy and
longitudinal momentum imbalance.  The dominating
backgrounds from hadronic $\tau$ decays and two-photon interactions
with hadronic final states are supressed to negligble levels.

\subsection{Event Shape Distributions}

The properties of hadronic events can be described by event shape
observables.  Event shape observables used for this analysis are
thrust (\thr) ~\cite{thrust1,thrust2}, heavy jet mass
(\mh)~\cite{jetmass}, wide and total jet broadening (\bw\ and
\bt)~\cite{nllabtbw}, C-Parameter (\cp)~\cite{parisi78,donoghue79,ert}
and the transition value between 2 and 3 jet
configurations~\cite{OPALPR003,komamiya90} defined by the Durham jet
algorithm (\ytwothree)~\cite{durham}.  Whenever we refer to a generic
event shape observable \thr, \mh, \bt, \bw, \cp\ or \ytwothree\ we use
the symbol $y$.

The event shape observables are calculated from selected charged
particle tracks and calorimeter clusters after correcting for double
counting of energy as described in~\cite{jader4}.  We compared the
data with the predictions from Monte Carlo simulations as described
above and found good agreement at all energy points.  Similar
observations were made in~\cite{jader4} for related observables.

\subsection{Corrections to the Data}
\label{detcorr}

The corrections to the data for limited experimental resolution,
acceptance and \bbbar\ background follows exactly
the treatment described in~\cite{jader4}.  Selected charged particle
tracks as well as electromagnetic clusters are used to calculate the
event shape observables.

From simulated events two different distributions are built: the {\em
detector-level} distribution and the {\em hadron-level distribution}.
The detector-level distributions are calculated exactly in the same
way as for data using measured charged particle tracks and calorimeter
clusters.  The hadron-level distributions use the true four-momenta of
the stable particles\footnote{All particles with lifetimes greater
than 300~ps are considered stable.} in events where the centre-of-mass
energy is reduced due to initial state radiation (ISR) by less than
0.15~GeV.  The bin-by-bin corrections for the data distributions are
derived from the ratio of hadron-level to detector-level distributions
for simulated events with u, d, s or c primary quarks.  Contributions
from B hadron decays bias the measurement of event shape observables
and therefore the expected contribution from $\epem\rightarrow\bbbar$
events is subtracted from the detector-level distributions before the
corrections are applied.
The simulations were optimised by OPAL to describe production
and decays of B hadrons~\cite{OPALPR141,OPALPR379}.   The good
description of our uncorrected data by the simulations confirms
that using the simulations to subtract the $\epem\rightarrow\bbbar$
background is justified.

In order to study systematic uncertainties the selection and
correction procedures are modified and the whole analysis is
repeated. The evaluation of the systematic uncertainties follows 
identically the procedure described in~\cite{jader4}.

\section{Determinations of \boldmath{\as}}

\subsection{QCD Calculations}
\label{qcdprediction}

The distributions of the event shape observables are predicted by \oaaa\ 
(NNLO) perturbative QCD calculations~\cite{gehrmannderidder07b}:
\begin{equation}
  \frac{1}{\sigma}\frac{d\sigma}{dy}_{\oaaa} 
    = \frac{dA}{dy}\ash + \frac{dB}{dy}\ash^2 + \frac{dC}{dy}\ash^3
\label{NNLOcalc}        
\end{equation}
with $\ash=\as(\mu)/(2\pi)$.  Equation~(\ref{NNLOcalc}) is shown for
renormalisation scale $\mu=Q$, where $Q$ is the physical scale usually
identified with the cms energy \roots\ for hadron production in \epem\
annihilation.  The coefficient distributions for leading order (LO)
$dA/dy$, next-to-leading order (NLO) $dB/dy$ and NNLO $dC/dy$ were
kindly provided by the authors of~\cite{gehrmannderidder07b}.
In~\cite{weinzierl08} a problem at small values of $y$ with the NNLO
terms calculated in~\cite{gehrmannderidder07b} was shown, but it does
not affect the kinematic regions selected in our fits.  The
normalisation to the total hadronic cross section and the terms
generated by variation of the renormalisation scale parameter
$\xmu=\mu/Q$ are implemented according to~\cite{gehrmannderidder07b}.
The prediction in equation~(\ref{NNLOcalc}) may be combined with
resummed NLLA calculations~\cite{gehrmann08} using the $\ln
R$-matching scheme; we refer to these predictions as NNLO+NLLA.  The
$\ln R$-matching procedure ensures that in the combination of the
fixed order NNLO calculations and the resummed NLLA calculations no
double counting of common terms occurs.  The NNLO+NLLA predictions are
here compared with experimental data for the first time.

The theoretical predictions provide distributions at the level of
quarks and gluons, the so-called {\em parton-level}.  The
distributions calculated using the final state partons after
termination of the parton showering in the models are also said to be
at the {\em parton-level}.  To compare the QCD predictions with
measured hadron-level event shape distributions the predictions are
corrected for hadronisation effects.  These corrections are obtained
by calculating in the Monte Carlo models the ratio of the cumulative
distributions at hadron-level and parton-level.  The corrections are
applied to the cumulative prediction $R(y)=\int_0^y 1/\sigma
d\sigma/dy'dy'$ as in~\cite{OPALPR158}.

It was shown in~\cite{dissertori07} that the event shape observable
distributions derived from the parton-level of the Monte Carlo
generators are described reasonably well by the NNLO calculation in
their fit ranges.  We compared the parton-level predictions of the
Monte Carlo generators with the QCD predictions in NNLO or NNLO+NLLA
with $\asmz=0.118$ and $\xmu=0.5, 1.0, 2.0$ at $\roots=14$, 22, 35 and
44~GeV.  We study the quantity $r(y)_{theory,MC}=d\sigma/dy_{theory} /
d\sigma/dy_{MC}-1$.  In addition we compute the corresponding
quantities $r(y)_{MC_i,MC_j}$ for any pair $i,j$ of Monte Carlo
predictions and $r(y)_{\xmu=0.5;1;2,\xmu=0.5;1;2}$ for theory
predictions with different renormalisation scale values.  The largest
values of the $\mathrm{abs}(r(y)_{MC_i,MC_j})$ and the
$\mathrm{abs}(r(y)_{\xmu=0.5;1;2,\xmu=0.5;1;2})$ at each $y$ are added
in quadrature to define the uncertainty $\Delta r(y)$ of
$r(y)_{theory,MC}$.  The average values $\bar{r}$ of the
$\mathrm{abs}(r(y))$ over the fit ranges (see below) are taken as a
measure of the consistency between between theory and Monte Carlo
predictions.  The ratios of $\bar{r}_{theory,MC}$ with the average
error $\Delta r(y)$ are generally smaller than or about equal to unity
and reach values of about two for \cp\ at $\sqrt{s}= 44$~GeV.  The
model dependence of the hadronisation correction and the
renormalisation scale dependence of the theory will be studied as
systematic uncertainties below. Our studies show that systematic
uncertainties introduced by discrepancies between the theory
predictions and the Monte Carlo parton-level predictions will be
covered by the combined hadronisation and theory systematic
variations.

\subsection{Measurements of \boldmath{\as}}

The strong coupling \as\ is determined by a \chisq\ fit to each of the
measured event shape distributions at the hadron-level, i.e.\
corrected for experimental effects.  A \chisq\ value is calculated at
each cms energy:
\begin{equation}
  \chisq= \sum_{i,j}^{n} (d_i-t_i(\as)) (V^{-1})_{ij} (d_j-t_j(\as))
\end{equation}
where $i,j$ count the bins within the fit range of the event shape
distribution, $d_i$ is the measured value in the $i$th bin, 
$t_i(\as)$ is the QCD prediction for the $i$th bin corrected for
hadronisation effects, and $V_{ij}$ is the covariance matrix of the
$d_i$.  The final prediction is obtained by integrating the QCD
predictions in equation~(\ref{NNLOcalc}) over the bin width after
application of the hadronisation correction as explained above in
section~\ref{qcdprediction}.  The \chisq\ value is minimised with
respect to \as\ while the renormalisation scale factor is set to
$\xmu=1$.  

The evolution of the strong coupling \as\ as a function of the
renormalisation scale is implemented in three loops as shown
in~\cite{kluth06}.  Since the evolution of \as\ in the cms energy
range considered here does not involve the crossing of flavour
thresholds it does not introduce significant uncertainties.  In order
to quantify the uncertainty from the evolution procedure we evolve
$\asmz=0.118$ from \mz\ to 14~GeV in three loops and two loops
and find a relative difference of 0.1\%.

In order to take the correlations between different bins into account
the covariance matrix $V_{ij}$ is computed following the approach
described in~\cite{OPALPR404}:
\begin{eqnarray}
V_{ij} & = & \sum_k \frac{\partial P_i}{\partial N_k}
                \frac{\partial P_j}{\partial N_k} N_k \\ \nonumber
       & = & \frac{1}{N^4} \sum_k \alpha_k^2 N_k
         \left(N\delta_{ik} - \widetilde{N}_i\right)
         \left(N\delta_{jk} - \widetilde{N}_j\right) \;.
\end{eqnarray}
$N_i$ is the number of data events, $\widetilde{N}_i=
\alpha_i(N_i-b_i)$ the number of events after subtraction of
background $b_i$ from \bbbar\ events and multiplying by a correction
$\alpha_i$ for detector effects, $P_i$ is the normalised hadron-level
distribution at bin $i$ and $N=\sum_k \widetilde{N}_k$.

The fit ranges are determined by several considerations.  We require
the leading log terms to be less than unity, because we also use a
fixed order expansion without resummation of log-enhanced terms, see
e.g.~\cite{ellis96}, and for consistency we use the same fit ranges in
the NNLO and NNLO+NLLA analyses.  The leading log term of $dA/dy$ is
$\ln y/y$ and we require $\ash\ln y/y<1$ for $\asmz=0.118$,
$14\le\roots\le200$~GeV and $y>0.1$.  The upper limit is given by the
requirement that all three orders of the NNLO calculations contribute,
i.e.\ the fit range extends to the kinematic limit of the LO
coefficients $dA/dy$.  The fit range for \mh\ should be compared with
the related fit range for \thr\ after squaring, because
$\mh^2\sim\thr$ in LO.  The fit ranges for
\cp\ and \thr\ are related by a factor of $(\ln6)/6\simeq0.3$.  The
resulting fit ranges are shown in table~\ref{fitranges}.

\begin{table}[htb!]
\caption{Fit ranges at all cms energies}
\label{fitranges}
\begin{tabular}{ ccc }
\hline\noalign{\smallskip}
\thr & \mh & \bt \\
0.10--0.27 & 0.26--0.50 & 0.16--0.30 \\
\noalign{\smallskip}\hline\noalign{\smallskip}
\bw & \cp & \ytwothree \\
0.10--0.23 & 0.34-0.72 & 0.01--0.20 \\
\noalign{\smallskip}\hline
\end{tabular}
\end{table}

The detector corrections (see section~\ref{detcorr}) are generally
$\pm20$\% or less within the fit ranges.  The hadronisation correction
factors are maximally 3.5 for \bt\ at $\roots=14$~GeV but are
generally smaller than 2 for the larger cms energy points.  The
hadronisation corrections are smallest and have the least variations
over the fit ranges for \ytwothree.

The evaluation of the systematic errors of the \as\ measurements
takes into account experimental effects, the hadronisation 
correction procedure and uncertainties of the theory.  The
three sources of systematic uncertainty are added in quadrature
to the statistical error taken from the fits to obtain
the total errors.  Below we describe how we find the systematic
uncertainties:
\begin{description}

\item[Experimental Uncertainties] The analysis is repeated with
  slightly varied event and track selection cuts and a systematic
  uncertainty from variation of the fit ranges is
  studied~\cite{jader4}.  The cross section used in the subtraction of
  \bbbar\ events is varied by $\pm5\%$ which takes account of possible
  differences in the efficiency determination using the simulations of
  $\epem\rightarrow\qqbar$ (q=u,d,s,c) and $\epem\rightarrow\bbbar$
  events.  For each experimental variation the value of \as\ is
  determined and compared to the central (default) value.  The
  quadratic sum of the differences and the fitrange uncertainty is
  taken as the experimental systematic uncertainty.

\item[Hadronisation] For the default analysis, PYTHIA is used to
  estimate the corrections originating from hadronisation effects
  (section~\ref{qcdprediction}).  As a systematic variation HERWIG and
  ARIADNE are used to evaluate the effects of hadronisation.  The larger
  of the deviations is taken as systematic hadronisation uncertainty.
  It was observed in~\cite{jones03,pedrophd} that systematic
  uncertainties between the PYTHIA, HERWIG and ARIADNE models are
  generally much larger than systematic uncertainties from varying the
  parameters of a given model.

\item[Theoretical Uncertainties] The theoretical prediction of event
  shape observables is a finite power series in \as.  The
  uncertainties originating from missing higher order terms are
  assessed by changing the renormalisation scale factor to $\xmu=0.5$
  and $\xmu=2.0$.  The larger deviation from the default value of \as\
  is taken as the systematic uncertainty.

\end{description}

\subsection{Results from NNLO Fits}
\label{sec_NLLOresults}

\begin{figure}[htb!]
\includegraphics[width=1.\columnwidth]{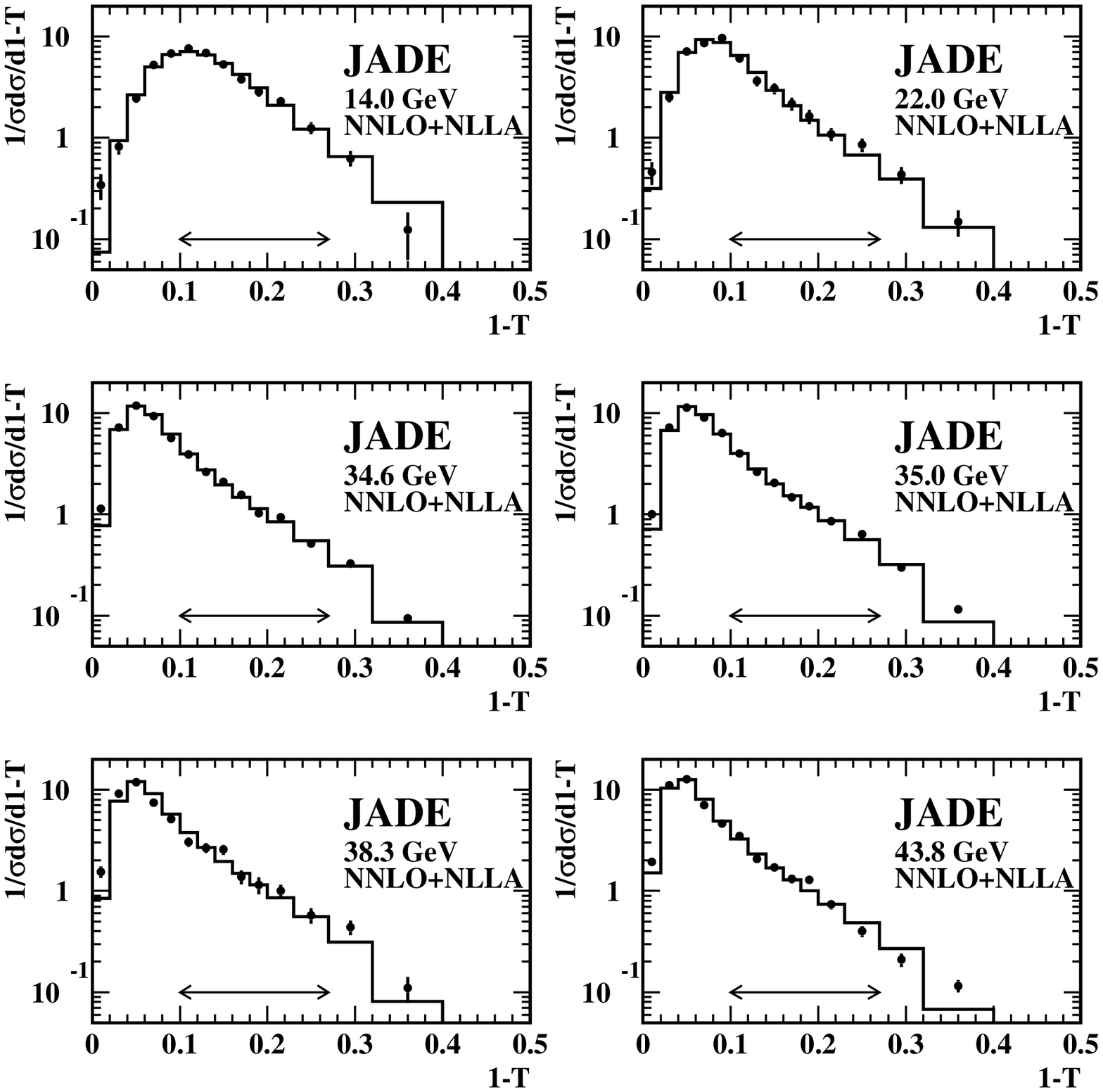} 
\caption{The plots show as points with statistical error bars the
\thr\ distributions at hadron level at $\roots=14$ to 43.8~GeV. Some
error bars are smaller than the data points.  Superimposed as
histograms are the NNLO+NLLA predictions combined with hadronisation
effects using the corresponding fit results for $\as(\roots)$ shown in
table~\ref{asresultsnnlonlla}.  The arrows indicate the fit ranges.}
\label{esdistributions}
\end{figure}

The results of the NNLO fits are summarised in
table~\ref{asresultsnnlo}.  In figure~\ref{esdistributions} the \thr\
event shape distributions together with the NNLO fit results for the
six energy points are shown.  The \chisqd\ values are between 0.7 for
$\roots=14$~GeV and 2.5 for $\roots=34.6$~GeV.  The fit results for
the other event shape observables return similar results with
$0.3<\chisqd<3.8$.  We note that at cms energies where we have big
data samples the \chisqd\ value tend to be larger.  The \chisqd\
values are based on statistical errors only while the combined
experimental and hadronisation uncertainties are at least a factor of
two larger than the statistical errors leading to a reduction of
\chisqd\ by a factor of at least 4 if these uncertainties were taken
into account in the fits.  We conclude that there is no significant
disagreement between the event shape data and the QCD fits.

The results at each cms energy are remarkably consistent with each
other, we find root-mean-square (rms) values for $\as(\roots)$ between
0.003 at 44~GeV and 0.008 at 22~GeV.  The hadronisation uncertainties
at 14~GeV dominate the total errors (except for \mh\ and \cp), because
at 14~GeV the hadronisation corrections are largest.  The statistical
errors are sizeable at $\roots=14$, 22, 38 and 44~GeV, where there is
only limited statistics and quite small at $\roots=34.6$ and 35~GeV
where we have large data samples.  The experimental uncertainties
depend somewhat on the cms energy with smaller values at higher
\roots\ where we have larger data samples.

The hadronisation uncertainties for \cp\ are the largest except at
$\roots=14$~GeV.  This as been observed
before~\cite{jadec,pedrophd,OPALPR425}.

\begin{table*}[htb!]
\caption{Results of NNLO fits to event shape observable distributions
at the JADE cms energies.}
\label{asresultsnnlo}
\begin{tabular}{ ccrrrrrc }
\hline\noalign{\smallskip}
\roots\ [GeV] & Obs. & $\as(\roots)$ & $\pm$stat. & $\pm$exp. & $\pm$had. & $\pm$theo. & \chisqd \\
\noalign{\smallskip}\hline\noalign{\smallskip}
14.0 & \thr &  0.1587 &  0.0098 &  0.0213 &  0.0366 &  0.0100 & $ 4.0/ 6$ \\
14.0 & \mh &  0.1759 &  0.0080 &  0.0133 &  0.0093 &  0.0099 & $ 9.0/ 5$ \\
14.0 & \bt &  0.1687 &  0.0086 &  0.0098 &  0.0337 &  0.0132 & $ 1.6/ 5$ \\
14.0 & \bw &  0.1730 &  0.0053 &  0.0088 &  0.0188 &  0.0088 & $ 3.8/ 5$ \\
14.0 & \cp &  0.1583 &  0.0150 &  0.0169 &  0.0113 &  0.0089 & $ 9.2/ 5$ \\
14.0 & \ytwothree &  0.1671 &  0.0039 &  0.0054 &  0.0101 &  0.0063 & $11.5/ 8$ \\
\noalign{\smallskip}\hline\noalign{\smallskip}
22.0 & \thr &  0.1410 &  0.0075 &  0.0054 &  0.0195 &  0.0070 & $ 6.8/ 6$ \\
22.0 & \mh &  0.1555 &  0.0070 &  0.0105 &  0.0064 &  0.0061 & $ 5.3/ 5$ \\
22.0 & \bt &  0.1399 &  0.0070 &  0.0050 &  0.0178 &  0.0076 & $ 4.0/ 5$ \\
22.0 & \bw &  0.1551 &  0.0046 &  0.0046 &  0.0106 &  0.0060 & $10.2/ 5$ \\
22.0 & \cp &  0.1385 &  0.0084 &  0.0073 &  0.0332 &  0.0062 & $ 3.7/ 5$ \\
22.0 & \ytwothree &  0.1545 &  0.0033 &  0.0025 &  0.0084 &  0.0049 & $10.6/ 8$ \\
\noalign{\smallskip}\hline\noalign{\smallskip}
34.6 & \thr &  0.1396 &  0.0017 &  0.0040 &  0.0100 &  0.0069 & $14.8/ 6$ \\
34.6 & \mh &  0.1477 &  0.0017 &  0.0070 &  0.0025 &  0.0053 & $18.9/ 5$ \\
34.6 & \bt &  0.1392 &  0.0016 &  0.0025 &  0.0063 &  0.0076 & $ 6.4/ 5$ \\
34.6 & \bw &  0.1457 &  0.0013 &  0.0047 &  0.0049 &  0.0049 & $ 3.5/ 5$ \\
34.6 & \cp &  0.1374 &  0.0017 &  0.0041 &  0.0126 &  0.0062 & $12.7/ 5$ \\
34.6 & \ytwothree &  0.1404 &  0.0009 &  0.0012 &  0.0061 &  0.0035 & $14.3/ 8$ \\
\noalign{\smallskip}\hline\noalign{\smallskip}
35.0 & \thr &  0.1430 &  0.0014 &  0.0040 &  0.0094 &  0.0074 & $14.0/ 6$ \\
35.0 & \mh &  0.1532 &  0.0014 &  0.0065 &  0.0023 &  0.0059 & $ 7.0/ 5$ \\
35.0 & \bt &  0.1432 &  0.0013 &  0.0042 &  0.0064 &  0.0082 & $13.8/ 5$ \\
35.0 & \bw &  0.1496 &  0.0011 &  0.0063 &  0.0046 &  0.0054 & $10.7/ 5$ \\
35.0 & \cp &  0.1427 &  0.0014 &  0.0036 &  0.0118 &  0.0069 & $ 9.3/ 5$ \\
35.0 & \ytwothree &  0.1451 &  0.0008 &  0.0021 &  0.0061 &  0.0039 & $15.0/ 8$ \\
\noalign{\smallskip}\hline\noalign{\smallskip}
38.3 & \thr &  0.1427 &  0.0049 &  0.0064 &  0.0081 &  0.0073 & $11.3/ 6$ \\
38.3 & \mh &  0.1578 &  0.0048 &  0.0067 &  0.0022 &  0.0066 & $ 2.0/ 5$ \\
38.3 & \bt &  0.1447 &  0.0045 &  0.0068 &  0.0048 &  0.0085 & $ 1.8/ 5$ \\
38.3 & \bw &  0.1488 &  0.0039 &  0.0069 &  0.0034 &  0.0053 & $ 3.1/ 5$ \\
38.3 & \cp &  0.1369 &  0.0049 &  0.0036 &  0.0101 &  0.0061 & $ 9.7/ 5$ \\
38.3 & \ytwothree &  0.1380 &  0.0028 &  0.0038 &  0.0071 &  0.0032 & $23.5/ 8$ \\
\noalign{\smallskip}\hline\noalign{\smallskip}
43.8 & \thr &  0.1341 &  0.0029 &  0.0034 &  0.0057 &  0.0061 & $11.3/ 6$ \\
43.8 & \mh &  0.1403 &  0.0029 &  0.0062 &  0.0014 &  0.0046 & $11.5/ 5$ \\
43.8 & \bt &  0.1312 &  0.0027 &  0.0027 &  0.0043 &  0.0064 & $ 9.8/ 5$ \\
43.8 & \bw &  0.1373 &  0.0023 &  0.0050 &  0.0028 &  0.0041 & $10.7/ 5$ \\
43.8 & \cp &  0.1342 &  0.0028 &  0.0042 &  0.0083 &  0.0058 & $ 4.3/ 5$ \\
43.8 & \ytwothree &  0.1333 &  0.0017 &  0.0026 &  0.0054 &  0.0029 & $21.0/ 8$ \\
\noalign{\smallskip}\hline
\end{tabular}
\end{table*}

\subsection{Results from NNLO+NLLA Fits}

The results of the NNLO+NLLA fits are given in
table~\ref{asresultsnnlonlla}.  The values of \chisqd\ are slightly
smaller than the corresponding values of the NNLO fits in most cases,
indicating a somewhat better consistency of the NNLO+NLLA calculations
with the fitted data points.  The rms values of $\as(\roots)$ are
between 0.007 at $\roots=22$~GeV and 0.003 at $\roots=44$~GeV, i.e.\
the scatter of individual results is essentially the same as for the
NNLO analysis.  The pattern of statistical errors and experimental and
hadronisation uncertainties is the same as for the NNLO fits discussed
above.  Compared with the NNLO analysis the theoretical uncertainties
are reduced by $10-20$\% and the values of \as\ are lower by 4\% on
average.  The hadronisation uncertainties of the NNLO+NLLA fits are
also smaller in most cases.  The difference in \as\ between NNLO and
NNLO+NLLA calculations is smaller than the difference in \as\ between
NLO and NLO+NLLA calculations as expected in~\cite{gehrmann08}.  As
discussed above in section~\ref{sec_NLLOresults} for NNLO fits the
sometimes large \chisqd\ values can be explained by the small
statistical errors in some data sets.

\begin{table*}[htb!]
\caption{Results of NNLO+NLLA fits to event shape observable distributions
at the JADE cms energies.}
\label{asresultsnnlonlla}
\begin{tabular}{ ccrrrrrc }
\hline\noalign{\smallskip}
\roots\ [GeV] & Obs. & $\as(\roots)$ & $\pm$stat. & $\pm$exp. & $\pm$had. & $\pm$theo. & \chisqd \\
\noalign{\smallskip}\hline\noalign{\smallskip}
14.0 & \thr &  0.1543 &  0.0092 &  0.0197 &  0.0362 &  0.0089 & $ 4.3/ 6$ \\
14.0 & \mh &  0.1641 &  0.0065 &  0.0105 &  0.0126 &  0.0095 & $ 9.2/ 5$ \\
14.0 & \bt &  0.1620 &  0.0078 &  0.0083 &  0.0343 &  0.0122 & $ 1.5/ 5$ \\
14.0 & \bw &  0.1540 &  0.0038 &  0.0060 &  0.0157 &  0.0072 & $ 1.4/ 5$ \\
14.0 & \cp &  0.1466 &  0.0128 &  0.0151 &  0.0131 &  0.0063 & $11.1/ 5$ \\
14.0 & \ytwothree &  0.1661 &  0.0038 &  0.0057 &  0.0129 &  0.0060 & $ 8.0/ 8$ \\
\noalign{\smallskip}\hline\noalign{\smallskip}
22.0 & \thr &  0.1383 &  0.0070 &  0.0051 &  0.0185 &  0.0061 & $ 6.1/ 6$ \\
22.0 & \mh &  0.1464 &  0.0058 &  0.0089 &  0.0040 &  0.0049 & $ 4.8/ 5$ \\
22.0 & \bt &  0.1350 &  0.0064 &  0.0044 &  0.0159 &  0.0062 & $ 3.4/ 5$ \\
22.0 & \bw &  0.1408 &  0.0034 &  0.0035 &  0.0074 &  0.0047 & $ 7.2/ 5$ \\
22.0 & \cp &  0.1337 &  0.0078 &  0.0064 &  0.0323 &  0.0061 & $ 4.3/ 5$ \\
22.0 & \ytwothree &  0.1538 &  0.0033 &  0.0022 &  0.0090 &  0.0045 & $ 8.5/ 8$ \\
\noalign{\smallskip}\hline\noalign{\smallskip}
34.6 & \thr &  0.1365 &  0.0016 &  0.0036 &  0.0092 &  0.0057 & $13.7/ 6$ \\
34.6 & \mh &  0.1399 &  0.0014 &  0.0055 &  0.0019 &  0.0041 & $15.9/ 5$ \\
34.6 & \bt &  0.1338 &  0.0014 &  0.0021 &  0.0055 &  0.0058 & $ 6.3/ 5$ \\
34.6 & \bw &  0.1332 &  0.0010 &  0.0034 &  0.0037 &  0.0035 & $ 6.2/ 5$ \\
34.6 & \cp &  0.1326 &  0.0016 &  0.0037 &  0.0111 &  0.0061 & $ 6.1/ 5$ \\
34.6 & \ytwothree &  0.1401 &  0.0009 &  0.0011 &  0.0059 &  0.0030 & $ 5.5/ 8$ \\
\noalign{\smallskip}\hline\noalign{\smallskip}
35.0 & \thr &  0.1399 &  0.0013 &  0.0037 &  0.0086 &  0.0062 & $10.5/ 6$ \\
35.0 & \mh &  0.1444 &  0.0012 &  0.0050 &  0.0015 &  0.0047 & $ 8.1/ 5$ \\
35.0 & \bt &  0.1376 &  0.0012 &  0.0038 &  0.0055 &  0.0065 & $11.9/ 5$ \\
35.0 & \bw &  0.1363 &  0.0008 &  0.0048 &  0.0035 &  0.0040 & $12.8/ 5$ \\
35.0 & \cp &  0.1373 &  0.0013 &  0.0032 &  0.0104 &  0.0068 & $ 4.6/ 5$ \\
35.0 & \ytwothree &  0.1447 &  0.0008 &  0.0022 &  0.0059 &  0.0035 & $ 9.6/ 8$ \\
\noalign{\smallskip}\hline\noalign{\smallskip}
38.3 & \thr &  0.1400 &  0.0046 &  0.0060 &  0.0073 &  0.0064 & $10.2/ 6$ \\
38.3 & \mh &  0.1484 &  0.0040 &  0.0050 &  0.0015 &  0.0056 & $ 1.9/ 5$ \\
38.3 & \bt &  0.1390 &  0.0041 &  0.0060 &  0.0043 &  0.0068 & $ 1.3/ 5$ \\
38.3 & \bw &  0.1360 &  0.0030 &  0.0052 &  0.0026 &  0.0042 & $ 2.5/ 5$ \\
38.3 & \cp &  0.1329 &  0.0045 &  0.0034 &  0.0085 &  0.0064 & $ 7.4/ 5$ \\
38.3 & \ytwothree &  0.1385 &  0.0028 &  0.0040 &  0.0063 &  0.0034 & $18.1/ 8$ \\
\noalign{\smallskip}\hline\noalign{\smallskip}
43.8 & \thr &  0.1313 &  0.0027 &  0.0031 &  0.0052 &  0.0048 & $11.6/ 6$ \\
43.8 & \mh &  0.1337 &  0.0025 &  0.0049 &  0.0011 &  0.0034 & $10.8/ 5$ \\
43.8 & \bt &  0.1265 &  0.0025 &  0.0023 &  0.0039 &  0.0047 & $10.0/ 5$ \\
43.8 & \bw &  0.1269 &  0.0018 &  0.0037 &  0.0022 &  0.0027 & $11.1/ 5$ \\
43.8 & \cp &  0.1295 &  0.0025 &  0.0037 &  0.0072 &  0.0054 & $ 3.5/ 5$ \\
43.8 & \ytwothree &  0.1331 &  0.0017 &  0.0024 &  0.0050 &  0.0024 & $18.6/ 8$ \\
\noalign{\smallskip}\hline
\end{tabular}
\end{table*}

\subsection{Combination of Results}

\begin{figure}[htb!]
\includegraphics[width=1.\columnwidth]{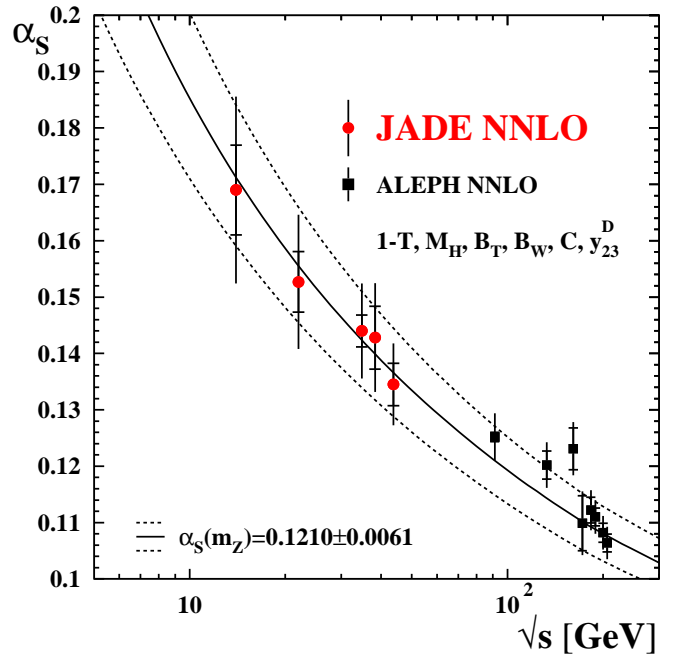} 
\caption{The values for \as\ at the JADE energy points.  The inner
error bars correspond to the combined statistical and experimental
errors and the outer error bars show the total errors.  The results
from $\roots=34.6$ and 35~GeV have been combined for clarity.  The
full and dashed lines indicate the result from our JADE NNLO analysis
as shown on the figure.  The results from the NNLO analysis of ALEPH
data~\cite{dissertori07} are shown as well.}
\label{asvscme}
\end{figure}

The results obtained at each energy point for the six event shape
observables are combined using error weighted averaging as
in~\cite{jones04,OPALPR404,aleph265,kluth06}.  The statistical
correlations between the six event shape observables are estimated at
each energy point from fits to hadron-level distributions derived from
50 statistically independent Monte Carlo samples.  The experimental
uncertainties are determined assuming that the smaller of a pair of
correlated experimental errors gives the size of the fully correlated
error (partial correlation).  The hadronisation and theory systematic
uncertainties are found by repeating the combination with changed
input values, i.e.\ using a different hadronisation model or a
different value of \xmu.  The results are given in
table~\ref{asecmcomb} and shown for the NNLO analysis in
figure~\ref{asvscme}, because this allows a direct
comparison with the results of the NNLO analysis of ALEPH
event shape data~\cite{dissertori07}.

The statistical uncertainties of the combined results are reduced as
expected.  The systematic uncertainties of the combined results tend
to be close to the best values from individual observables, because
the systematic uncertainties are not completely correlated and because
observables with smaller uncertainties have larger weights in the
combination procedure.

\begin{table}[htb!]
\caption{Combined values of $\as(\roots)$ at the JADE cms energies
from NNLO (upper section) and NNLO+NLLA (lower section) analyses
together with the statistical, experimental, hadronisation and theory
errors.}
\label{asecmcomb}
\begin{tabular}{ crrrrr }
\hline\noalign{\smallskip}
\roots\ [GeV] & $\as(\roots)$ & $\pm$stat. & $\pm$exp. & $\pm$had. & $\pm$theo. \\
\noalign{\smallskip}\hline\noalign{\smallskip}
14.0 & 0.1690 & 0.0046 & 0.0065 & 0.0124 & 0.0076 \\
22.0 & 0.1527 & 0.0040 & 0.0036 & 0.0090 & 0.0056 \\
34.6 & 0.1420 & 0.0012 & 0.0025 & 0.0058 & 0.0050 \\
35.0 & 0.1463 & 0.0010 & 0.0032 & 0.0059 & 0.0055 \\
38.3 & 0.1428 & 0.0033 & 0.0045 & 0.0060 & 0.0051 \\
43.8 & 0.1345 & 0.0021 & 0.0031 & 0.0043 & 0.0045 \\
\noalign{\smallskip}\hline\noalign{\smallskip}
14.0 & 0.1605 & 0.0044 & 0.0065 & 0.0148 & 0.0073 \\
22.0 & 0.1456 & 0.0036 & 0.0033 & 0.0077 & 0.0048 \\
34.6 & 0.1367 & 0.0011 & 0.0023 & 0.0046 & 0.0040 \\
35.0 & 0.1412 & 0.0009 & 0.0032 & 0.0049 & 0.0047 \\
38.3 & 0.1388 & 0.0030 & 0.0043 & 0.0042 & 0.0048 \\
43.8 & 0.1297 & 0.0019 & 0.0028 & 0.0033 & 0.0034 \\
\noalign{\smallskip}\hline
\end{tabular}
\end{table}

A combination of the combined results at the six JADE energy points
shown in table~\ref{asecmcomb} after running to a common reference
scale \mz\ using the combination procedure described above results in
\begin{eqnarray}
\label{equresultnnlo}
  \asmz & = 0.1210 & \pm 0.0007\stat\pm 0.0021\expt \\ \nonumber
        &          & \pm 0.0044\had\pm 0.0036\theo
\end{eqnarray}
($\asmz = 0.1210\pm0.0061$) for the NNLO analysis and 
\begin{eqnarray}
\label{equresultnnlonlla}
  \asmz & = 0.1172 & \pm 0.0006\stat\pm 0.0020\expt \\ \nonumber
        &          & \pm 0.0035\had\pm 0.0030\theo
\end{eqnarray}
($\asmz=0.1172\pm0.0051$) for the NNLO+NLLA analysis.  The NNLO+NLLA
result has smaller hadronisation and theory uncertainties compared
with the values in the NNLA analysis.  We choose the latter result
from NNLO+NLLA fits as our final result, because it is based on the
most complete theory predictions and it has smaller theory
uncertainties.  It is consistent with the world average of
$\asmz=0.119\pm0.001$~\cite{bethke06}, the recent NNLO analysis of
event shape data from the ALEPH experiment
$\asmz=0.1240\pm0.0033$~\cite{dissertori07} as well as with the
related average of $\asmz=0.120\pm0.005$ from the analyses of the LEP
experiments using NLO+NLLA QCD predictions~\cite{kluth06}.  The total
error for \asmz\ of 4\% is among the most precise determinations of
\asmz\ currently available.

\begin{figure}[htb!]
\begin{tabular}[tbp]{cc}
\includegraphics[width=1.\columnwidth]{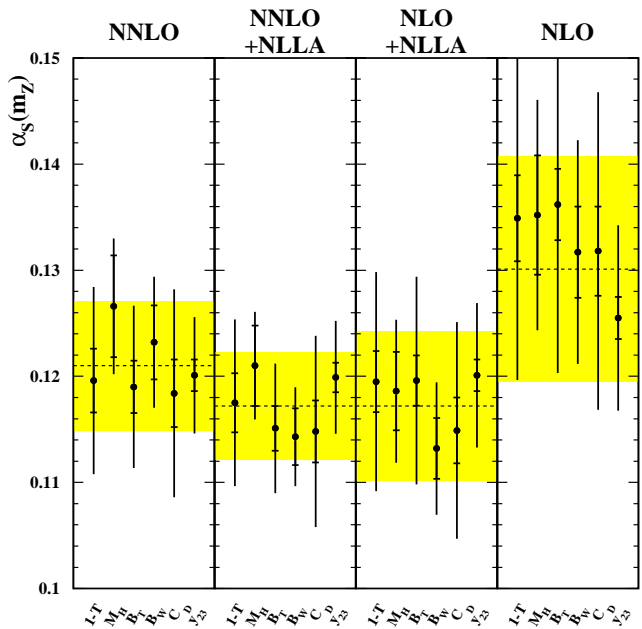} 
\end{tabular}
\caption{The combined results for \asmz\ for each type of
analysis as indicated on the figure.  The shaded bands and
dashed lines show the combined values of \asmz\ with total
errors.  The inner and outer error bars show the combined
statistical and experimental and the total errors. }
\label{asobsscatter}
\end{figure}

After running the fit results for $\as(\roots)$ for each observable to
the common reference scale \mz\ we combine the results for a given
observable to a single value.  We use the same method as above and
obtain the results for \asmz\ shown in table~\ref{asvarcomb}.  The rms
values of the results for \asmz\ are 0.0029 for the NNLO analysis and
0.0026 for the NNLO+NLLA analysis; both values are consistent with the
errors of the corresponding combined results shown in
equations~(\ref{equresultnnlo}) and~(\ref{equresultnnlonlla}).
Figure~\ref{asobsscatter} shows the combined results of \asmz\ for
each observable together with results from alternative analyses
discussed below.  Combining the combined results for each observable
or combining all individual results after evolution to the common
scale \mz\ yields results consistent with
equation~(\ref{equresultnnlonlla}) within $\Delta\asmz=0.0004$ and
the uncertainties also agree.

The hadronisation uncertainty of \mh\ at each energy point and in the
combinations shown in table~\ref{asvarcomb} is the smallest.  We have
repeated the combinations without \mh\ and found results for \asmz\
consistent within 0.6\% with our main results with hadronisation
uncertainties increased by 14\% (NNLO) or 20\% (NNLO+NLLA).

\begin{table}[htb!]
\caption{Combined values of \asmz\ for each observable from NNLO
(upper section) and NNLO+NLLA (lower section) analyses together with
the statistical, experimental, hadronisation and theory errors.}
\label{asvarcomb}
\begin{tabular}{ crrrrr }
\hline\noalign{\smallskip}
Obs. & \asmz & $\pm$stat. & $\pm$exp. & $\pm$had. & $\pm$theo. \\
\noalign{\smallskip}\hline\noalign{\smallskip}
\thr & 0.1196 & 0.0011 & 0.0028 & 0.0067 & 0.0049 \\
\mh & 0.1266 & 0.0009 & 0.0047 & 0.0014 & 0.0040 \\
\bt & 0.1190 & 0.0009 & 0.0023 & 0.0047 & 0.0055 \\
\bw & 0.1232 & 0.0008 & 0.0034 & 0.0037 & 0.0035 \\
\cp & 0.1184 & 0.0013 & 0.0029 & 0.0081 & 0.0045 \\
\ytwothree & 0.1201 & 0.0005 & 0.0014 & 0.0046 & 0.0026 \\
\noalign{\smallskip}\hline\noalign{\smallskip}
\thr & 0.1175 & 0.0010 & 0.0026 & 0.0061 & 0.0041 \\
\mh & 0.1210 & 0.0008 & 0.0037 & 0.0011 & 0.0032 \\
\bt & 0.1151 & 0.0009 & 0.0019 & 0.0039 & 0.0042 \\
\bw & 0.1143 & 0.0006 & 0.0026 & 0.0028 & 0.0026 \\
\cp & 0.1148 & 0.0011 & 0.0027 & 0.0073 & 0.0044 \\
\ytwothree & 0.1199 & 0.0005 & 0.0013 & 0.0046 & 0.0023 \\
\noalign{\smallskip}\hline
\end{tabular}
\end{table}

In order to study the compatibility of our data with the QCD
prediction for the evolution of the strong coupling with cms energy we
repeat the combinations with or without evolution of the combined
results to the common scale.  We set the theory uncertainties to zero
since these uncertainties are highly correlated between energy points.
We conservatively assume the hadronisation uncertainties to be
partially correlated, because these uncertainties depend strongly on
the cms energy.  The \chisq\ probabilities of the averages for running
(not running) with NNLO+NLLA fits then become 0.39 ($9.9\cdot
10^{-3}$).  With the NNLO fits the \chisq\ probabilities for running
(not running) are 0.48 ($1.2\cdot 10^{-3}$).  We interpret this as
strong evidence for the dependence of the strong coupling on cms
energy as predicted by QCD from JADE data alone.

\subsection{Comparison with NLO and NLO+NLLA}
\label{compNLONLLA}

For a comparison of our results with previous \as\ measurements the
fits to the event shape distributions are repeated with NLO
predictions and with NLO predictions combined with resummed NLLA with
the modified $\ln R$-matching scheme (NLO+NLLA), both with $\xmu=1$.
The NLO+NLLA predictions with the modified $\ln R$-matching scheme
were the standard of the final analysis of the LEP
experiments~\cite{l3290,aleph265,OPALPR404,delphi327}.  The fit ranges
as well as the procedures for evaluation of the systematic
uncertainties are identical to the ones in our NNLO and NNLO+NLLA
analyses.

The combination of the fits using NLO predictions returns 
$\asmz= 0.1301\pm 0.0009\stat\pm 0.0029\expt\pm 0.0054\had\pm 0.0086\theo$,
the fits using combined NLO+ NLLA predictions yield 
$\asmz= 0.1172\pm 0.0007\stat\pm 0.0022\expt\pm 0.0039\had\pm 0.0054\theo$
and these results are shown in figure~\ref{asobsscatter}.
The result obtained with the NLO+NLLA prediction is consistent with
our NNLO and NNLO+NLLA analyses, but the theory uncertainties are
larger by about 60\%.  The analysis using NLO predictions gives
theoretical uncertainties larger by a factor of 2.6 and the value for
\asmz\ is larger compared to the NNLO or NNLO+NLLA results.  It has
been observed previously that values for \as\ from NLO analysis with
$\xmu=1$ are large in comparison with most other
analyses~\cite{OPALPR054}.  Both the NLO+NLLA or NNLO+NLLA analyses
yield a smaller value of \asmz\ compared to the respective NLO or NNLO
results.  The difference between NNLO+NLLA and NNLO is smaller than
the difference between NLO+NLLA and NLO, since a larger part of the
NLLA terms is included in the NNLO predictions.

\subsection{Renormalisation Scale Dependence}

\begin{figure}[htb!]
\includegraphics[width=0.75\columnwidth]{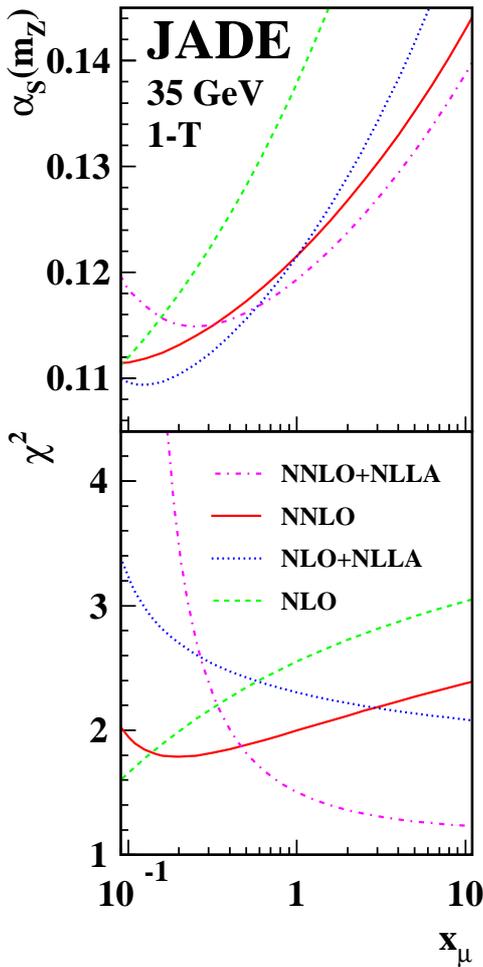} 
\caption{The plot shows the result of \asmz\ and $\chi^{2}$/d.o.f.
of the fit to the thrust event shape distribution for $\roots=35$~GeV.}
\label{FigureScaleDep}
\end{figure}

The theoretical uncertainty due to missing higher order terms is
evaluated by setting the renormalisation scale parameter \xmu\ to 0.5
or 2.0.  In order to assess the dependence of \as\ on the
renormalisation scale the fits are repeated using NNLO, NNLO+NLLA, NLO
and NLO+NLLA predictions with $0.1<\xmu<10$.  The strong coupling
\as\ as well as the \chisqd\ as a function of \xmu\ for \thr\ at
$\roots=35$~GeV are shown in figure~\ref{FigureScaleDep}.  The
\chisqd\ curves for the NLO+NLLA and NNLO+NLLA fits show no local
minimum in the \xmu\ range studied.  The \asmz\ values from NLO
predictions are the largest for $\xmu>0.2$.  The \asmz\ values using
NLO+NLLA and NNLO calculation almost cross at the natural choice of
the renormalisation scale $\xmu=1$ while the \asmz\ value from the
NNLO+NLLA fit is slightly lower.  The NLLA terms at $\xmu=1$ are
almost identical to the \oaaa-terms in the NNLO calculation.  A
similar behaviour can be observed for \bt\ and \ytwothree.
The slopes of the \asmz\ curves of the NNLO and NNLO+NLLA
fits around the default choice $\xmu=1$ are smaller than the slopes for
the NLO and NLO+NLLA fits leading to the decreased theoretical
uncertainties in our analyses.

\section{Conclusion}

In this paper we present measurements of the strong coupling \as\
using event shape observable distributions at cms energies
$14<\roots<44$~GeV.  To determine \as\ fits using NNLO and combined
NNLO+NLLA predictions were used.  Combining the results from NNLO+NLLA
fits to the six event shape observables \thr, \mh, \bw, \bt, \cp\ and
\ytwothree at the six JADE energy points returns 
$\asmz= 0.1172\pm 0.0006\stat\pm 0.0020\expt\pm 0.0035\had\pm 0.0030\theo$,
with a total error on \asmz\ of 4\%.  The investigation of the
renormalisation scale dependence of \asmz\ shows a reduced dependence
on \xmu\ when NNLO or NNLO+NLLA predictions are used, compared to
analyses with NLO or NLO+NLLA predictions.  The more complete NNLO or
NNLO+NLLA QCD predictions thus lead to smaller theoretical
uncertainties in our analysis.  The combined results for \as\ at each
cms energy are consistent with the running of \as\ as predicted by QCD
and exclude absence of running with a confidence level of 99\%.

%\bibliographystyle{ephja}
%\bibliography{jouabrv,opal,aleph,delphi,l3,bibdbs}

\end{document}